\documentclass[11pt,twoside,onecolumn]{article}
\usepackage[dvips]{epsfig}
\newcommand{\be}{\begin{eqnarray}}
\newcommand{\ee}{\end{eqnarray}}
\newcommand{\dbar}{\mbox{{\rm d\hspace{-2truemm}$^-$}}}
\title{\bf Effective Action and Thermodynamics of Radiating Shells
in General Relativity}
\author{G.L. Alberghi\thanks{e-mail: alberghi@bo.infn.it},\
R. Casadio\thanks{e-mail: casadio@bo.infn.it}
\ and G. Venturi\thanks{e-mail: armitage@bo.infn.it}\\
 \\
{\em Dipartimento di Fisica, Universit\`a di Bologna} \\
{\em and} \\
{\em Istituto Nazionale di Fisica Nucleare,
Sezione di Bologna, Italy}}
\begin{document}
%
%
\maketitle
\begin{abstract}
An effective action is obtained for the area and mass aspect of a thin
shell of radiating self-gravitating matter.
On following a mini-superspace approach, the geometry of the embedding
space-time is not dynamical but fixed to be either Minkowski or
Schwarzschild inside the shell and Vaidya in the external space
filled with radiation.
The Euler-Lagrange equations of motion are discussed
and shown to entail the expected invariance of the effective
Lagrangian under time-reparametrization.
They are equivalent to the usual junction equations and suggest a
macroscopic quasi-static thermodynamic description.
\end{abstract}
%
\pagestyle{plain}
\raggedbottom
\setcounter{page}{1}
\section{Introduction}
\setcounter{equation}{0}
The Einstein field equations for a general distribution of matter are
a formidable challenge and the study of the collapse of self-gravitating
compact objects is a very hard task to which much effort has been
dedicated since the birth of General Relativity
(see, {\em e.g.}, Ref.~\cite{mtw}).
One of the first papers in this field \cite{oppenheimer} showed
that, when gravity overcomes all other forces, a sphere of matter
collapses under its own weight into a point-like singularity.
This opened up a whole line of investigation about the nature of
such a singularity and the way it forms.
\par
A state corresponding to a point-like singularity
would violate Heisenberg's uncertainty principle, therefore
such a problem makes physical sense in a region for which gravity
can be treated at the (semi)classical level (see {\em e.g.}
\cite{bv,bfv,ebo} and Refs. therein).
Provided the conditions for such an approximation hold, one is left with
two types of difficulties which conspire against the achievement
of a definitive answer:
firstly, one would like to consider a realistic model for the collapsing
matter and, secondly, one has to face the intrinsic non-linearity
of the Einstein field equations.
The former aspect includes the quantum nature of matter as described by
the standard model of elementary particles with all of its own
intricacies.
The latter difficulty becomes greater along with the complications of the
matter model but, in turn, provides the most interesting features.
Thus one has to find a sensible compromise between reality and practical
solvability of the equations.
\par
A great deal of simplification usually follows from global space-time
symmetry.
The natural framework for the study of gravitational collapse is
isotropy with respect to a point.
Of course this rules out rotating objects, such as stars and other
astrophysical systems, and freezes many of the degrees of freedom of
gravity (Birkhoff's theorem forbids the emission of gravitational
waves \cite{mtw}).
However this does not render the field equations trivial and some of
the strong field effects certainly survive in this approximation
(see, {\em e.g.}, Ref.~\cite{choptuik}).
\par
An example of manageable isotropic distributions is given by dust fluid
models, such as the Tolman-Bondi space-time \cite{tolman}, which can be used
to represent spheres of pressureless matter and cosmological models.
Such space-times develop caustics, where the energy density of the fluid
diverges, because the geodesics followed by dust particles
cross for generic initial conditions.
The dynamics can be investigated by approximating the continuous
distribution of matter by a discrete medium of nested homogeneous
time-like shells of small but finite thickness.
When dust particles are confined by their own weight to roughly within
a few Compton wavelengths and the latter is negligible with respect
to any other ``macroscopic'' length in the model (to wit,
the square root of the area of the shell and its Schwarzschild
radius \cite{acvv}), one can take the limit of infinitely small
thickness ({\em thin shell limit}) and treat each shell as a singular
hypersurface generated by a $\delta$-function
term in the matter stress-energy tensor \cite{israel,hk}.
Then the basic problems are the study the dynamics of one shell
and the collisions between two nearby shells
(an issue we do not touch upon here).
\par
The method for treating singular hypersurfaces in General Relativity
was formulated in Ref.~\cite{israel} as following from the Einstein
field equations.
It amounts to the Lanczos junction equations between the embedding
metrics and entails the conservation of the total stress-energy.
The equations of motion of thin time-like shells have then
been derived for all spherically symmetric embedding four-geometries
\cite{lake}, including the vacuum \cite{israel} and
bubbles of non-vanishing cosmological constant \cite{guth}.
It is remarkable and physically sound that, in non-radiating cases,
the trajectory of the area of the shell is determined
by the junction equations once one has fixed the two external
metrics and the equation of state for the density and the surface
tension (the radial pressure must vanish).
The special case of radiating shells was studied in Ref.~\cite{pim}
for unpolarized radiation of relatively high frequency which behaves
as null dust and gives rise to a Vaidya external geometry
\cite{vaidya,israel2}.
This is indeed a good approximation when the radiation
wavelength is much smaller than any other ``macroscopic'' length
in the model (see, {\em e.g.} \cite{bicak} and Refs. therein) and
is reasonably consistent within our approach,
since the same condition on the Compton wavelength of the matter in
the shell is required by the thin shell limit \cite{acvv}.
\par
For the purpose of developing a quantum theory, knowledge
of the equations of motion is not sufficient, instead one needs
an (effective) action for the shell degrees of freedom.
The conceptual starting point is thus the Einstein-Hilbert action
rather than Einstein field equations.
In Ref.~\cite{hk} both metric and matter degrees of freedom
were kept dynamical and the general form of the action was given
for a barotropic fluid with step- or $\delta$-function
discontinuity.
One then implements the symmetry of the system in order to reduce
the action to a manageable effective form.
The literature on this topic treats essentially two
approaches for the shell:
\begin{description}
\item[{\em Canonical approach.}]
In Ref.~\cite{h1}, the embedding empty space-time is foliated
into spatial sections of constant time according to the ADM prescription
\cite{adm}, with lapse and shift functions in the four-metric.
The Einstein-Hilbert Lagrangian from Ref.~\cite{hk} is then integrated
over the spatial sections by making use of the properties of a spherical
vacuum \cite{kuchar}.
This leads to a canonical effective action for the canonical variables
of the system which include the inner and outer Schwarzschild times on
the shell as reminders of the geometrodynamics of the embedding
space-time.
A set of constraints ensures the invariance of the action under
reparametrization.
The shell matter Lagrangian used in \cite{h1} is not in the form of a
field theory and does not give rise to any canonical variable.
\item[{\em Mini-superspace approach.}]
In Refs.~\cite{guth,balbinot} the embedding metrics are chosen
{\em a priori\/} to be specific solutions of Einstein field equations
and are not dynamical.
The corresponding contribution to the Einstein-Hilbert Lagrangian
can then be integrated over a convenient space-time volume
and expressed in terms of metric variables of the shell world-volume
\cite{balbinot}.
This yields an effective action which is invariant under
reparametrization of the time on the shell and is equivalent to
the canonical one in domains of the phase space where some of the
constraints are solved classically \cite{h2}.
Thus one expects that the mini-superspace approach gives a limited
information on the quantum theory of geometry fully addressed in the
canonical approach, but does not limit the possibility of taking into
account the quantum nature of matter in a semi-classical context,
as was later done in Ref.~\cite{acvv}.
\end{description}
\par
In the present paper we illustrate the derivation of an effective
action for a
time-like shell which can emit unpolarized high frequency radiation.
Since the basic aim is to establish a starting point for the study of
semi-classical effects induced by the quantum evolution of the matter
in the shell (along the lines of Ref.~\cite{acvv}), the mini-superspace
approach will suffice.
The dynamical meaning of such an action will just be to generate the
time evolution of the shell degrees of freedom, which will be identified
with the area and mass aspect of the shell.
However, we shall also see that this approach, unlike the standard
junction prescriptions, allows a simple interpretation of
the evolution equations and time-reparametrization invariance in terms
of the thermodynamics of the shell.
\par
The plan of the paper is as follows: in Section~\ref{action}
we start from the general Einstein-Hilbert action as given in
Ref.~\cite{hk} and simplify it in order to derive a mini-superspace
effective action.
In Section~\ref{motion} we write the Euler-Lagrange equations of
motion, two of which are equivalent to the junction equations
and compare with previous approaches.
In section~\ref{thermo} we illustrate a thermodynamic
interpretation and in Section~\ref{conc} we finally comment our results.
We shall follow the sign convention of Ref.~\cite{mtw}
(see also Appendix~\ref{scalar}) with Greek indices $\mu,\nu,\ldots$
labeling space-time four-coordinates and Latin indices $i,j,\ldots$
for the coordinates of a three-dimensional sub-manifold;
$\kappa=16\,\pi\,G$, with $G$ the Newton constant.
\section{Derivation of the action}
\setcounter{equation}{0}
\label{action}
The space-time $\Omega$ we are considering is parted into two regions,
$\Omega^\pm$, separated by the shell world-volume $\Sigma$.
$\Omega^-$ inside the shell is devoid of matter and $\Omega^+$
possibly contains out-flowing radiation.
The corresponding Einstein-Hilbert action, from Ref.~\cite{hk},
is thus
\be
S_{EH}&=&{1\over\kappa}\,\int_{\Omega^-} d^4x\,\sqrt{-g}\,{\cal R}
\nonumber \\
&&+{2\over\kappa}\,\int_\Sigma d^3x\,\sqrt{h}\,
\left[{\cal K}\right]^+_-
-\int_\Sigma d^3x\,\sqrt{h}\,\rho
\nonumber \\
&&+{1\over\kappa}\,\int_{\Omega^+} d^4x\,\sqrt{-g}\,{\cal R}
+\int_{\Omega^+} d^4x\,\sqrt{-g}\,{\cal L}_{rad}
\ ,
\label{SEH}
\ee
where $g$ is the determinant of the four-dimensional metric, $h$
the determinant of the three-dimensional metric on $\Sigma$,
${\cal R}$ is the scalar curvature, ${\cal K}$ the trace of the extrinsic
curvature of $\Sigma$, $\rho$ the energy density of the shell and
${\cal L}_{rad}$ the radiation Lagrangian density.
We have also introduced the notation $\left[F\right]^+_-$
for the difference between the limiting (on the shell)
values of a function $F$ computed in $\Omega^+$ and $\Omega^-$.
\par
The above action is the starting point for our derivation.
First of all we observe that the spherical symmetry of the system
makes it possible to introduce three global coordinates
$(r,\theta,\phi)$, with $r>0$ the radial coordinate of a sphere of
area $4\,\pi\,r^2$, $\theta\in(0,\pi)$ and $\phi\in(0,2\,\pi)$
the usual angular coordinates.
The next step is to decide which one of the two approaches
outlined in the introduction is better suited for the present
problem.
Since a formulation of the kind given in Ref.~\cite{kuchar}
is not available at present for a radiation-filled space-time
\cite{bp}, the canonical approach is not viable.
Therefore, we follow Ref.~\cite{balbinot} and assume the Einstein field
equations are satisfied inside the space-time volumes $\Omega^\pm$
\cite{Kuc}.
\par
The above assumption fixes the form of the embedding metrics.
In particular, we observe that $\Omega$ can be
naturally parted into three regions (see Fig.~\ref{adm}):
\begin{figure}
\centerline{
\epsfysize=250pt\epsfbox{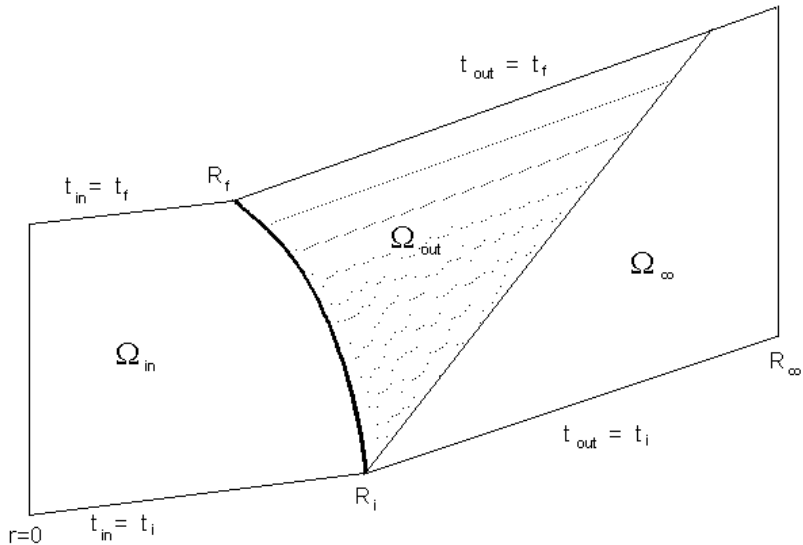}\\}
\caption{The space-time of the collapsing shell with the ADM foliation.}
\label{adm}
\end{figure}
\par\noindent
{\em i)} the inner empty space $\Omega_{in}=\Omega^-$, whose geometry
can be either Minkowski or Schwarzschild.
Let $t_{in}$ be the (Schwarzschild) time coordinate in $\Omega_{in}$,
then the corresponding metric can be written
\be
ds^2_{in}=-a_{in}\,dt_{in}^2+a_{in}^{-1}\,dr^2+r^2\,d\Omega^2
\ ,
\label{schw_in}
\ee
where $d\Omega^2=d\theta^2+\sin^2\theta\,d\phi^2$ is the usual line
element on the two-sphere and $a_{in}=1-2\,M_{in}/r$.
The constant $M_{in}$ is positive for Schwarzschild and zero for
Minkowski;
\par\noindent
{\em ii)} an outer space $\Omega_{out}$ which can be filled with
radiation.
If $t_{out}$ denotes the (Schwarzschild) time in $\Omega_{out}$,
the metric, which is Vaidya \cite{vaidya}, takes the diagonal form
\be
ds^2_{out}=b^2_{out}\,a_{out}^{-1}\,dt_{out}^2+a_{out}^{-1}\,dr^2
+r^2\,d\Omega^2
\ ,
\label{v_diag}
\ee
where $b_{out}=\partial_t m/\partial_r m$, $a_{out}=1-2\,m/r$.
The {\em mass aspect}, $m=m(t_{out},r)$, equals the total energy
included inside the sphere of area $4\,\pi\,r^2$ at time $t_{out}$
\cite{mtw};
\par\noindent
{\em iii)} an empty space $\Omega_\infty$ prior to any emission
of radiation with Schwarzschild metric and mass parameter equal
to the total ADM mass $M_\infty$ of the system,
\be
ds^2_\infty=-a_\infty\,dt^2_\infty+a_\infty^{-1}\,dr^2+ r^2\,d\Omega^2
\ .
\label{schw_inf}
\ee
where $a_\infty=1-{2\,M_\infty/ r}$.
\par
In this frame, the shell trajectory is given by $r=R(\tau)$ and
$t_{in/out}=T_{in/out}(\tau)$, where $\tau$ is the arbitrary time
variable on the shell world-volume $\Sigma$ with three-metric $h_{ij}$
given by
\be
ds^2_\Sigma=-N^2\,d\tau^2+R^2\,d\Omega^2
\ ,
\label{3_g}
\ee
$N=N(\tau)$ being the shell lapse function.
The full stress-energy tensor of the matter in the system contains
two parts,
\begin{eqnarray}
{\cal T}_{\mu\nu}={\cal S}_{ij}\,\delta(r-R)
+{\cal T}_{\mu\nu}^{rad}\,\theta(r-R)
\ ,
\label{T}
\end{eqnarray}
where $\delta$ is the Dirac $\delta$-function and $\theta$ the
step-function.
The source term,
\begin{eqnarray}
{\cal S}_{ij}={\rm diag}\,
\left[N^2\,\rho,-R^2\,P,-R^2\,\sin^2\theta\,P\right]
\ ,
\label{S}
\end{eqnarray}
is the three-dimensional stress-energy tensor of a fluid with density
$\rho$ and (surface) tension $P$ and ${\cal T}_{\mu\nu}^{rad}$ is the
stress-energy tensor of the out-flowing null radiation.
\par
The mass aspect can be extended to all values of $r>0$, does not
decrease for increasing $r$ and must be continuous in $\Omega$ except
at the shell radius, thus
\be
&&\lim_{r\to R^-} m(T_{in},r)=M_{in}
\nonumber \\
&&\lim_{r\to R^+} m(T_{out},r)=M_{out}\ge M_{in}
\nonumber \\
&&\lim_{r\to \infty} m(t_\infty,r)=M_\infty\ge M_{out}
\ ,
\label{m_match}
\ee
where $M_{out}=M_{out}(\tau)$ is the mass aspect at the (outer) shell
radius and equals the total energy of the shell plus $M_{in}$.
Other matching conditions will be implemented in the following
and, as in Eq.~(\ref{m_match}), we shall use capital letters for the
restriction (or limit) to the shell position of functions of
the space-time coordinates which are denoted by the corresponding
small letters, {\em e.g.}, $A_{in}=a_{in}(R,T)$.
We shall usually drop the subscripts $in/out$ whenever this does not
cause any confusion ({\em e.g.}, $t=t_{in}$ for $r<R$) and
total derivatives with respect to $\tau$ are denoted by a dot.
\par
Since the metrics above solve the Einstein equations with the source
(\ref{T}), the volume contributions in the action
(\ref{SEH}) must vanish identically.
Further, the matching between $\Omega_{out}$ and $\Omega_\infty$ is
smooth, provided the radial coordinate $R_s$ of the border between
the two regions satisfies \cite{vaidya}
\be
{dR_s\over dt_\infty}=1-{2\,M_\infty\over R_s}
\ ,
\ee
where $M_\infty$ equals $M_{out}$ at the time $\tau$ at which the
emission starts, and no boundary terms arise at the surface $r=R_s$.
The reduced Einstein-Hilbert action can then be written
\be
S_{EH}^{red}&=&{8\,\pi\over\kappa}\,\int_{\tau_i}^{\tau_f} d\tau\,N\,R^2\,
\left[{\cal K}\right]^{out}_{in}
-4\,\pi\,\int_{\tau_i}^{\tau_f} d\tau\,N\,R^2\,\rho
+S_B
\nonumber \\
&\equiv&S_G^{shell}+S_M^{shell}+S_B
\ ,
\label{EH}
\ee
where $S_G^{shell}$ is the surface gravitational action of the shell,
$S_B$ contains all surface contributions at the border of the space-time
$\Omega$ (including those which are usually introduced
to cancel second derivatives of the dynamical variables)
and $S_M^{shell}$ is the shell matter action,
\be
S_M^{shell}=-4\,\pi\,\int_{\tau_i}^{\tau_f} d\tau\,N\,R^2\,
\rho
\ ,
\label{SMsh}
\ee
where we assume $\rho$ does not depend on $N$ and its dependence on
the other two shell degrees of freedom $R$ and $M_{out}$ will be clarified
in Section~\ref{motion}.
\par
At this point one can foresee a technical difficulty in evaluating
surface terms at $t$ constant because the region $\Omega_{out}$ is not
homogeneous (see Fig.~\ref{adm}).
In fact, the mass aspect $m$ is constant along the lines
of the outgoing flow of radiation and one can introduce a null
coordinate $u=u_{out}(t_{out},r)$ such that out-going null
geodesics are given by $u$ constant with four-velocity
$k^\mu\sim(0,1,0,0)$ and one has \begin{eqnarray}
{\cal T}^{rad}_{\mu\nu}=
-{4\over\kappa}\,{1\over r^2}\,{dm\over du}\,\delta_\mu^u
\,\delta_\nu^u
\ .
\end{eqnarray}
The mass aspect $m=m(u)$ does not depend on $r$ now,
and the metric is written (for $r>R$)
\be
ds^2_{out}=-a_{out}\,du^2-2\,du\,dr+r^2\,d\Omega^2
\ .
\label{vaidya}
\ee
Such a $u_{out}$ is defined also for $M_{out}$ constant, in which case
it coincides with a null Eddington-Finkelstein coordinate for the
Schwarzschild space-time \cite{mtw} (see also Appendix~\ref{null}).
Integration over $r$ at time $t_{out}$ constant would then involve
integrating functions of $m$ and its derivatives, that is the knowledge
of $M(T)$ and $R(T)$, for all $t_i\le T\le t_{out}$.
However, although the differential relation between $(t_{out},r)$ and
$(u_{out},r)$ is known,
\be
du=-a^{-1}_{out}\,\left(b_{out}\,dt_{out}+dr\right)
\ ,
\label{du}
\ee
$u=u_{out}(t_{out},r)$ cannot be written explicitly unless the
functions $M_{out}(\tau)$ and $R(\tau)$ are fixed once and for all.
At the same time the trajectory of the shell is to be derived from
the sought effective action and cannot be given {\em a priori}.
This renders explicit integration over slices of constant $t$
inconvenient.
\par
\begin{figure}
\centerline{
\epsfysize=270pt\epsfbox{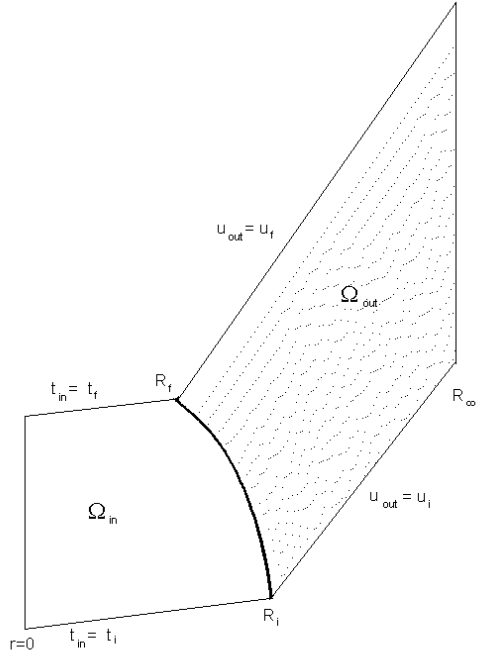}\\}
\caption{The space-time $\Omega_{in}\cup\Omega_{out}$ with the
``mixed'' foliation used for the computation of the effective action.}
\label{mixed}
\end{figure}

The above argument suggests one define a ``mixed'' foliation, where
volume sections are defined by $t$ constant for $r<R$ and $u$
constant for $r>R$ (see Appendix~\ref{null} for an equivalent
overall null foliation).
Hence we shall integrate the Einstein-Hilbert action in the four-volume
$\Omega=\Omega_{in}\cup\Omega_{out}$ displayed in Fig.~\ref{mixed}.
Such volume has a time-like boundary at $r=R_\infty$ and two ``mixed''
boundaries defined by $t_{in}=t_i$, $u_{out}=u_i$ and $t_{in}=t_f>t_i$,
$u_{out}=u_f$.
The shell trajectory is described equivalently by $r=R(\tau)$,
$t=T(\tau)$, with (fixed) end-points $R_i=R(\tau_i)$ at $T(\tau_i)=t_i$
and $R_f=R(\tau_f)$ at $T(\tau_f)=t_f$, in $\Omega_{in}$
and $r=R(\tau)$, $u=U(\tau)$ in $\Omega_{out}$, with end-points
at $U(\tau_i)=u_i$ and $U(\tau_f)=u_f$.
\par
The fact that all volume terms (the integrals over $\Omega^\pm$
in Eq.~(\ref{SEH})) do not contribute to the reduced action
(\ref{EH}) can now be checked explicitly.
The basic observation is to note that the scalar curvature for a metric
of the form (\ref{vaidya}) or (\ref{g_in}) is given by
(see Appendix~\ref{scalar})
\be
{\cal R}=
{2\over r^2}\,\left[1-{\partial\over\partial r}\left(a\,r
+{r^2\over 2}\,{\partial a\over \partial r}\right)\right]
\ ,
\label{Rv}
\ee
and vanishes identically once one substitutes in $a=a_{out}$ for
Vaidya or $a=a_{in}$ for Schwarzschild.
The corresponding volume action is thus
\be
{4\,\pi\over\kappa}\,\int_{t_i}^{t_f} dt\,\int_0^R r^2\,dr\,
{\cal R}_{in}+
{4\,\pi\over\kappa}\,\int_{u_i}^{u_f} du\,\int_R^{R_\infty} r^2\,dr\,
{\cal R}_{out}=0
\ ,
\label{SGext}
\ee
as expected.
One also recalls that the Lagrangian density for a field of null
dust is proportional to $k_\mu\,k^\mu$, where $k^\mu$ is the
fluid four-velocity, and thus vanishes along null geodesics.
\par
In order to compute $S_G^{shell}$
we introduce Gaussian normal coordinates $(\tau,\eta,\theta,\phi)$
near the shell such that the trajectory $R(\tau)$ is given by $\eta=0$
with $\eta>0$ for $r>R$.
Then, the jump in the components of the extrinsic curvature of the shell
world-volume are given by the relation \cite{mtw}
\be
\left[{\cal K}_{ij}\right]^{in}_{out}=
\lim\limits_{\epsilon\to 0^+}\left(
{1\over 2}\,\left.{\partial g_{ij}\over\partial\eta}
\right|_{\eta=-\epsilon}
-{1\over 2}\,\left.{\partial g_{ij}\over\partial\eta}
\right|_{\eta=+\epsilon}
\right)
\ .
\label{k}
\ee
In detail (see Appendix~\ref{gauss})
\be
{\cal K}_{\theta\theta}={{\cal K}_{\phi\phi}\over\sin^2\theta}=\left\{
\begin{array}{ll}
R\,\strut\displaystyle{\beta_{in}\over N} &\ \ \ \ \ \ \eta=-\epsilon
\\
 \\
R\,\strut\displaystyle{\beta_{out}\over N} &\ \ \ \ \ \ \eta=+\epsilon
\ ,
\end{array}\right.
\label{Kthth}
\ee
\be
{\cal K}_{\tau\tau}=\left\{\begin{array}{ll}
\strut\displaystyle{1\over \beta_{in}}\,\left[\dot R\,\dot N
-\strut\displaystyle{N^3\over 2}\,
\strut\displaystyle{\partial A_{in}\over\partial R}
-\ddot R\,N\right] &\ \ \ \ \ \ \eta=-\epsilon
\\
 \\
\strut\displaystyle{1\over \beta_{out}}\,\left[\dot R\,\dot N
-\strut\displaystyle{N^3\over 2}\,
\strut\displaystyle{\partial A_{out}\over\partial R}
-\ddot R\,N
-\strut\displaystyle{N^3\over 2}\,\strut\displaystyle
\left({\beta_{out}-\dot R\over A_{out}\,N}\right)^2\,
{\partial A_{out}\over \partial U}\right] &\ \ \ \ \ \ \eta=+\epsilon
\ ,
\end{array}\right.
\label{Ktt}
\ee
where
\be
\beta\equiv\sqrt{A\,N^2+\dot R^2}
\ .
\label{beta}
\ee
The above expressions give
\be
S_G^{shell}={8\,\pi\over\kappa}\,\int_{\tau_i}^{\tau_f} d\tau\,\left[
2\,\beta\,R+{R^2\over\beta}\,\left\{\ddot R-\dot R\,{\dot N\over N}
+{N^2\over 2}\,\left[{\partial A\over\partial R}
+\left({\beta-\dot R\over A\,N}\right)^2\,{\partial A\over \partial U}
\right]\right\}\right]^{in}_{out}
\ ,
\label{SGsh1}
\ee
where we used $\partial a_{in}/\partial u_{in}=0$ to write the limits
from the two regions in the same form.
\par
The integral in Eq.~(\ref{SGsh1}) contains a second derivative of
$R$ which should be eliminated by a suitable term in $S_B$.
Such term must be a total derivative so that it does not affect
the equations of motion and for the Schwarzschild metric it can be found
in Ref.~\cite{guth}.
The same kind of term works for Vaidya as well and is given by
\be
S_B^{(1)}=-{8\,\pi\over\kappa}\,\,\int_{\tau_i}^{\tau_f} d\tau\,
{d\over d\tau}\left[R^2\,\tanh^{-1}
\left({\dot R\over\beta}\right)\right]^{in}_{out}
\ .
\label{SB1}
\ee
Adding (\ref{SB1}) to (\ref{SGsh1}) gives
\be
S_G^{shell}+S_B^{(1)}=
{8\,\pi\over\kappa}\,\int_{\tau_i}^{\tau_f} d\tau\,\left[
2\,\beta\,R-2\,\dot R\,R\,\tanh^{-1}\left({\dot R\over\beta}\right)
+{M\,\beta\over A}-{\dot M\,R\over A}\right]^{in}_{out}
\ .
\label{SGsh}
\ee
We observe that $\beta_{in}/A_{in}=\dot T_{in}$ \cite{guth}
and the mass aspect $M_{in}$ is constant, therefore the
last two terms are irrelevant in $\Omega_{in}$ (being zero or
total derivatives).
In $\Omega_{out}$, however, they cannot be neglected.
\par
We shall now identify another surface term of dynamical relevance
which arises from the borders of $\Omega$.
We observe that the expression (\ref{Rv}) of the scalar curvature
is defined up to the addition of zero which can be written in the
form
\be
0={2\,\alpha\over r^2}\,{\partial\over\partial r}(r-a\,r)
+{2\,\gamma\over r^2}\,{\partial\over\partial r}\left(
{r^2\over 2}\,{\partial a\over \partial r}\right)
\ ,
\ee
where $\alpha$ and $\gamma$ are arbitrary constants.
Hence
\be
&&{4\,\pi\over\kappa}\,\int_{u_i}^{u_f} du\,\int_R^{R_\infty} r^2\,dr\,
{\cal R}=
\nonumber \\
&&{8\,\pi\over\kappa}\,\int_{u_i}^{u_f} du\,\int_R^{R_\infty} dr\,\left[
\left(1+\alpha\right)-
\left(1+\alpha\right)\,{\partial \over \partial r}\left(a\,r\right)
-\left(1-\gamma\right)\,{\partial \over \partial r}\left(
{r^2\over 2}\,{\partial a\over \partial r}\right)\right]
\ ,
\label{SGout}
\ee
from which one readily sees that there are no (non-trivial)
surface terms at the null borders $u=u_i$ and $u=u_f$ of $\Omega_{out}$.
It is shown in Ref.~\cite{guth} that no such terms arise along the
spatial borders $t=t_i$ and $t=t_f$ of $\Omega_{in}$.
Therefore, from (\ref{SGout}) one concludes that a surface term could
come only from the border at $r=R_\infty$ and should be compensated
by
\be
S_B^{(2)}&=&{8\,\pi\over\kappa}\,\int_{u_i}^{u_f} du\,
\left[(1+\alpha)\,a\,r
+(1-\gamma)\,{r^2\over 2}\,{\partial a\over \partial r}\right]_{R_\infty}
\nonumber \\
&=&-{8\,\pi\over\kappa}\,\int_{u_i}^{u_f} du\,\left[
\left(1+2\,\alpha+\gamma\right)\,M_{out}
-\left(1+\alpha\right)\,R_\infty\right]
\ .
\label{SB}
\ee
For $\alpha=1$ and $\gamma=0$ one obtains the usual contribution
(see Appendix~\ref{border}).
However we find it more useful to set $\alpha=-1$ and $\gamma=0$,
for which the integrand in Eq.~(\ref{SB}) with the relation
(\ref{dotU}) becomes
\be
S_B^{(2)}&=&{8\,\pi\over\kappa}\,\int_{\tau_i}^{\tau_f} d\tau\,
{\beta_{out}-\dot R\over A_{out}}\,M_{out}
\ ,
\label{SB2}
\ee
since this form cancels one of the terms in the integral (\ref{SGsh}).
\par
Putting together all the relevant pieces given in Eqs.~(\ref{SMsh}),
(\ref{SGsh}) and (\ref{SB2}) finally yields the effective action
\be
S_{eff}(N,R,M_{out})&=&{8\,\pi\over\kappa}\,\int_{\tau_i}^{\tau_f} d\tau\,
\left\{\left[2\,R\,\beta
-2\,R\,\dot R\,\tanh^{-1}\left({\dot R\over\beta}\right)
\right]_{out}^{in}
+{\dot M_{out}\,R\over A_{out}}
-{M_{out}\,\dot R\over A_{out}}
\right\}
\nonumber \\
&&-4\,\pi\,\int_{\tau_i}^{\tau_f} d\tau\,N\,R^2\,\rho
\nonumber \\
&\equiv&\int_{\tau_i}^{\tau_f} d\tau\,L_{eff}(N,R,M_{out})
\ ,
\label{Seff1}
\ee
for the variables $N$, $R$ and $M_{out}$.
It is worth noting that one recovers the effective action for a shell
which collapses without emission of radiation simply by setting
$\dot M_{out}=0$.
In fact the integral of $M_{out}\,\dot R/A_{out}$, which does not appear
in the action given in Ref.~\cite{balbinot}, depends only on the
end-points when $M_{out}$ is a constant and is then dynamically
irrelevant.
\par
In the next Section we shall vary $S_{eff}$ in order to check that
it generates the same equations of motion as following from the
junction conditions reviewed in appendix~\ref{junction}.
\section{Equations of motion}
\setcounter{equation}{0}
\label{motion}
In the following we shall obtain the Euler-Lagrange equations of
motion by varying the effective action with respect to the
$\tau$-dependent quantities appearing in the effective Lagrangian
with the end-points held fixed at $\tau_i$ and $\tau_f$.
Such variations are formally computed by acting on the Lagrangian
with the operator ($i=1,2,3$)
\be
{\delta\over\delta X^i}\equiv
{\partial\over\partial X^i}
-{d\over d\tau}\left({\partial\ \over\partial \dot X^i}\right)
\ ,
\label{delta}
\ee
with $X^i=(N,R,M_{out})$ and
$\delta X^i(\tau_i)=\delta X^i(\tau_f)=0$.
\par
To begin with, we note that it is sensible in the spirit of General
Relativity to assume that the shell proper energy $E$ does
not depend on the shell velocity $\dot R$.
Further, in order to keep a sufficient degree of generality, we
consider $E$ depending on both shell variables $R$
and $M_{out}$ (but not on $N$),
so that the physical shell energy is related to the
quantity $4\,\pi\,R^2\,\rho$ by
\be
E=4\,\pi\,R^2\,\rho(R,M_{out})
\ ,
\label{EF}
\ee
and agrees with the source in the $\theta\theta$-junction equation
(\ref{match1}).
\par
We next introduce the canonical momenta
$P_i\equiv(\partial L_{eff}/\partial \dot X^i)$, which are formally
given by
\be
&&P_N=0
\label{PN}
\\
&&P_R=-{8\,\pi\over\kappa}\,\left\{2\,R\,\left[
\tanh^{-1}\left({\dot R\over\beta}\right)\right]^{in}_{out}
+{M_{out}\over A_{out}}\right\}
\label{PR}
\\
&&
P_M={8\,\pi\over\kappa}\,{R\over A_{out}}
\label{PM}
\ .
\ee
The particular expressions (\ref{PN}) and (\ref{PM}) deeply affect
the nature of the dynamical system at hand, as can be seen from the
(symmetric) matrix
\be
W_{ij}&\equiv&{\partial L_{eff}\over\partial\dot X^i\,\partial\dot X^j}
={\partial P_i\over\partial\dot X^j}
\ ,
\ee
whose rank equals the number of canonical degrees of freedom with
${\rm Dim}[W]-{\rm Rank}[W]$ being the number of primary constraints.
Since
\be
{\rm Rank}[W]={\rm Rank}\left[{\rm diag}
\left(\begin{array}{ccc}
0\,, & \strut\displaystyle{\partial P_R\over\partial\dot R}\,,
& 0
\end{array}\right)\right]
=1
\ ,
\label{W}
\ee
the Lagrangian $L_{eff}$ contains two primary constraints and is said
to be {\em non-standard\/} \cite{sud}.
\par
Varying $S_{eff}$ with respect to the lapse function yields
\be
{\delta L_{eff}\over \delta N}={\partial L_{eff}\over \partial N}
={16\,\pi\over\kappa}\,{R\over N}\,\left[\beta\right]^{in}_{out}
-E
\ ,
\label{H}
\ee
where we made use of our previous assumption that $\rho$ does not
depend on $N$.
Upon setting this variation to zero we obtain the first primary
(Hamiltonian) constraint
\be
{H\over N}\equiv E
-{16\,\pi\over\kappa}\,{R\over N}\,\left[\beta\right]^{in}_{out}
=0
\ ,
\label{em1}
\ee
which is related to the time-reparametrization invariance of the shell
three-metric and is formally equal to the analogous constraint obtained in
Ref.~\cite{balbinot} for non-radiating shells.
In order to make contact with the notation in Appendix~\ref{junction}
we observe that, for $N=1$, one formally recovers the
$\theta\theta$-junction equation (\ref{match1}).
\par
Varying with respect to $R$ yields the equation of motion
\be
{\delta L_{eff}\over \delta R}=
{\partial L_{eff}\over \partial R}
-{d\over d\tau}\left({\partial L_{eff}\over \partial \dot R}\right)
=0
\ ,
\ee
that is
\be
{\delta\over\delta R}\left(N\,E\right)=
{16\,\pi\over\kappa}\,\left\{\left[\beta+{1\over\beta}\,\left(
R\,\ddot R+{N^2\,M\over R}-{R\,\dot N\,\dot R\over N}
\right)\right]^{in}_{out}
+{\dot M_{out}\over\beta_{out}}\,
{\beta_{out}-\dot R\over 1-2\,M_{out}/R}\right\}
\ .
\label{em2}
\ee
Upon defining
\be
P\equiv
{1\over 8\,\pi\,N\,R}\,{\delta\over\delta R}\left(N\,E\right)
={\delta E\over\delta{\cal A}}
\ ,
\label{Prho}
\ee
(${\cal A}=4\,\pi\,R^2$ is the area of the surface of the shell)
and setting $N=1$ we formally recover the $\tau\tau$-junction
equation (\ref{match2}).
\par
Now, unlike the non-radiating case, one must also consider varying
$M_{out}$, thus obtaining the second primary constraint which has no
interpretation as a junction condition,
\be
{\delta L_{eff}\over \delta M_{out}}&=&
{\partial L_{eff}\over \partial M}
-{d\over d\tau}\left({\partial L_{eff}\over \partial \dot M}\right)
\nonumber \\
&=&
{16\,\pi\over\kappa}\,{\beta_{out}-\dot R\over 1-2\,M_{out}/R}
-{\delta\over\delta M_{out}}\left(N\,E\right)=0
\ .
\label{em3}
\ee
\par
It was already obvious from Eq.~(\ref{PN}) and $W_{11}=0$ that the
lapse function $N$ is a Lagrange multiplier and can be assigned any
function of $\tau$.
Hence, from now on we work in the {\em proper time\/} gauge $N=1$, which
makes most of the equations look simpler.
The corresponding constraint (\ref{em1}) was in anticipation named
the Hamiltonian constraint since the quantity $N\,H$ is the canonical
Hamiltonian, as can be seen from the Hamiltonian form of the effective
Lagrangian,
\be
L_{eff}=P_N\,\dot N+P_R\,\dot R+P_M\,\dot M_{out}-N\,H
\ .
\ee
The Hamiltonian constraint must be preserved in time, which yields the
secondary constraint
\be
{d H\over d\tau}&=&{dE\over d\tau}-{\partial E\over\partial R}\,\dot R
-{\partial E\over\partial M_{out}}\,\dot M_{out}
\nonumber \\
&=&4\,\pi\,R^2\left[\dot\rho-2\,{\dot R\over R}\,\left(P-\rho\right)
-{4\over\kappa}\,{\dot M_{out}\over R}\,
{\beta_{out}-\dot R\over R-2\,M_{out}}\right]
=0
\ ,
\label{dH}
\ee
where we have used both Eqs.~(\ref{em2}) and (\ref{em3}) to show that,
in general, Eq.~(\ref{dH}) is trivially satisfied by our {\em ansatz}
$E=E(R,M_{out})$ in Eq.~(\ref{EF}).
\par
Since $W_{33}=0$, also the quantity $M_{out}$ is not a true dynamical
variable.
Further, the total time derivative of (\ref{em3}) vanishes in virtue of
$E=E(R,M_{out})$ on using Eqs.~(\ref{em3}) itself and (\ref{dH}),
therefore no new secondary constraint is generated.
This means that one can set $M_{out}=M_{out}(\tau)$ (any function
of $\tau$), which determines $E=E(R,M_{out})$ through the constraints
(\ref{em1}) and (\ref{em3}), and compute $R=R(\tau)$ with correspondingly
$R(\tau_i)=R_i$ by solving Eq.~(\ref{em2}).
This picture is well suited to describe a shell whose luminosity
curve in time is given by \cite{pim}
\be
{d Q\over d\tau}\equiv {16\,\pi\over\kappa}\,
{\beta_{out}-\dot R\over 1-2\,M_{out}/R}\,{d M_{out}\over d\tau}
\ .
\label{dQ}
\ee
The above quantity is negative for radiating shells (we required
$\dot U>0$, see Eq.~(\ref{dotU})) and diverges for the shell approaching
its Schwarzschild radius \cite{lind}, unless
\be
\dot M_{out}=\dot U\,{dM_{out}\over du}
\sim(R-2\,M_{out})
\ ,
\label{finite}
\ee
for $R\to 2\,M_{out}$.
This singular behaviour (and the corresponding bound on $\dot M_{out}$)
is due to the fact that the coordinates $(u,r)$ are related to the point
of view of a static observer (see Appendix~\ref{junction}) and is removed
by passing to Israel's coordinates \cite{israel2} (see also
Appendix~\ref{gauss}).
\par
However, one can alternatively choose the value of
$E=E(R(\tau),M_{out}(\tau))$ as an explicit function of $\tau$ and
determine $P$ from (\ref{em2}) and the luminosity from (\ref{em3}).
Then the trajectory $R=R(\tau)$ is obtained by imposing the Hamiltonian
constraint (\ref{em1}) for all times $\tau_i\le \tau\le \tau_f$.
This option is particularly useful if one wishes to impose only
initial conditions and then consider both the trajectory and the
luminosity of the shell as completely determined by the
specific interaction between the shell matter and the emitted
radiation encoded in the dependence of $E$ on $R$ and $M_{out}$.
\subsection{Comparison with other approaches}
In a previous approach \cite{guth,balbinot} the outer mass
is considered fixed, in which case it is not possible nor
necessary to consider the variation with respect to it.
Nontheless one may proceed in analogy with such an approach
and obtain a self-consistent, complete set of equations starting
from our action Eq.~(\ref{Seff1}).
\par
The first case one may consider is that for which the arbitrary
function $M_{out}(\tau)$ is taken to be a function of the shell
radius $M_{out}=M_{out}(R(\tau))$.
One then has that the two
terms to the right of eq. (\ref{Seff1}) are a total
derivative with respect to the proper time,
\be
\int_{\tau_i}^{\tau_f} d\tau\, \left({\dot M_{out}\,R\over A_{out}}
-{M_{out}\,\dot R\over A_{out}} \right)
=-\int_{\tau_i}^{\tau_f} d\tau\,\dot R\,{R^2\over 2}\,
\left({\partial\over\partial R}\,\log A_{out}\right)
\ ,
\ee
which is a boundary term and does not influence the equation
of motion.
Then the tension is found to be
\be
P={2\over\kappa}\,\left[{\beta\over R}+{\ddot R\over \beta}
+{1\over 2\,\beta}\,{dA\over dR}\right]^{in}_{out}
\ee
which is the analogue of eq. (\ref{em2}) and agrees with the
$\tau\tau$-junction equation.
\par
Using this expression and imposing the conservation of the
Hamiltonian constraint we obtain as anticipated a continuity
equation, which takes the form
\be
{d\rho\over dR}={2\over R}\,(P-\rho)
\ .
\ee
\par
Finally for the general case $ M_{out}(\tau)$ we note that
Eq.~(\ref{em3}), which was obtained by varying the action
with respect to $ M_{out}$, can also be arrived at by using
the expression for the pressure (that is the equation of motion
for $R$) and imposing the conservation of the Hamiltonian
constraint
\be
{dH\over d\tau}=0
\ .
\ee
It is clear, as we have mentioned after Eq.~(\ref{Seff1}),
that for $M_{out}$ constant we reproduce the usual results
\cite{guth,balbinot}.
\section{Thermodynamics}
\label{thermo}
Just as for the black hole case \cite{hawking}, one may attempt for the
case of the radiating shell a thermodynamic approach.
The first observation is that in thermodynamics one has ``quasi-static''
processes so that the equations of state are always satisfied.
This of course implies in our case that all time derivatives be
considered small so that matter on the shell always be in equilibrium
\cite{york}.
Let us illustrate such an approach (in this Section we denote the
quasi-static limit of previously defined quantities by the same letter
in italics style).
\par
In the shell one expects that the intensive coordinate be the
surface tension ${\cal P}$ and the extensive one be the area
${\cal A}$.
Thus, considering quasi-static processes for which
\be
\dot R^2\ll 1-{2\,M_{out}\over R}
\ ,
\ee
from Eq.~(\ref{em2}) one finds (again setting $N=1$)
\be
{\cal P}={2\over\kappa\,R}\,
\left[\sqrt{A}+{M\over R\,\sqrt{A}}\right]_{out}^{in}
\ .
\label{4.1}
\ee
Similarly, from Eq.~(\ref{em1}) one may identify
\be
{\cal E}={16\,\pi\over\kappa}\,R\,\left[\sqrt{A}\right]_{out}^{in}
\ ,
\label{4.2}
\ee
which will correspond to the shell internal energy.
We note that all quantities are a function of the (slowly)
time-dependent variables $M_{out}$ and $R$, thus it is clear that
$d{\cal E}$ (a change of the internal energy) is an exact differential
\cite{X} of $M_{out}$ and $R$, which is a statement of the first law of
thermodynamics \cite{zemansky}.
\par
One may now introduce a variation of the heat ${\cal Q}$,
\be
\dbar{\cal Q}&=&d{\cal E}-{\cal P}\,d{\cal A}
\nonumber \\
&=&d{\cal E}-8\,\pi\,R\,{\cal P}\,dR
\nonumber \\
&=&{\partial{\cal E}\over\partial M_{out}}\,dM_{out}
+\left({\partial{\cal E}\over\partial R}-
8\,\pi\,R\,{\cal P}\right)\,dR
\ ,
\ee
which on using Eqs.~(\ref{4.1}) and (\ref{4.2}) leads to
\be
\dbar{\cal Q}={16\,\pi\over\kappa}\,{dM_{out}\over
\sqrt{1-2\,M_{out}/R}}
\ .
\ee
This is in agreement with the quasi-static limit of Eq.~(\ref{em3})
and is related to the luminosity of the shell, Eq.~(\ref{dQ}).
At this point, in order to obtain the change in entropy, one must
introduce a temperature ${\cal T}$ which must be such that
$d{\cal S}=\dbar{\cal Q}/{\cal T}$ is an exact differential as a
statement of the second law of thermodynamics \cite{zemansky}.
This implies that
\be
{\partial\over\partial R}\,\left(
{\cal T}\,\sqrt{1-2\,M_{out}/R}\right)^{-1}=0
\ ,
\label{Teq}
\ee
which has a solution \cite{fpq}
\be
{\cal T}={\hbar\over 8\,\pi\,k_b\,M_{out}}\,
{1\over\sqrt{1-2\,M_{out}/R}}
\ ,
\label{Tyork}
\ee
where $k_b$ is the Boltzmann constant, and corresponds to the
desired equation of state connecting $M_{out}$, $R$ and ${\cal T}$.
The entropy change in this case is then given by
\be
d{\cal S}=\left({16\,\pi\,k_b\over\hbar\,\kappa}\right)\,8\,\pi\,
M_{out}\,dM_{out}
\ ,
\ee
and for $\kappa=16\,\pi$, $\hbar=k_b=1$ one has (omitting an
arbitrary constant)
\be
{\cal S}=4\,\pi\,M_{out}^2={1\over 4}\,(\hbox{area of horizon})
\ ,
\ee
which is the usual result for a black hole and it is clear that a
collapse for which $\dot M_{out}=0$ is adiabatic (``isentropic'').
\par
Obviously, given our expressions for the entropy, surface tension and
internal energy, one may actually evaluate the various thermodynamic
potentials.
Let us just end by determining the specific heat at constant area.
It will be given by
\be
C_{\cal A}={\cal T}\,\left({\partial{\cal S}\over\partial{\cal T}}
\right)_{\cal A}
=-{1\over 8\,\pi\,{\cal T}^2}\,\left(1-{3\,M_{out}\over R}\right)^{-1}
\ ,
\ee
which for $R>3\,M_{out}$ is negative as expected.
\section{Conclusions}
\setcounter{equation}{0}
\label{conc}
In this paper we have analyzed the dynamics of radiating shells in
General Relativity by deriving a mini-superspace effective action
for the area and mass aspect of the shell.
Besides proving the dynamical equivalence of this approach with the
usual treatment via junction equations, we have introduced a temperature
(equation of state) and obtained a thermodynamic, quasi-static
description for the evolution of the shell.
\par
The effective action in itself is useful for developing the quantum
theory and we hope to use it along the lines of Ref.~\cite{acvv}.
We also think that the remarks in the Introduction and the analysis
of the equations of motion carried on in the previous Section make
clear that description given for radiating shells is a general
framework which can be used to study a wide variety of cases of
physical interest.
\appendix
\setcounter{equation}{0}
\section{Overall null foliation}
\label{null}
Instead of the ``mixed'' foliation used in the text, one might
introduce a null Eddington-Finkelstein coordinate \cite{mtw}
$du_{in}=dt_{in}-a_{in}^{-1}\,dr$ such that
\be
ds^2_{in}=-a_{in}\,du_{in}^2-2\,du_{in}\,dr+r^2\,d\Omega^2
\ ,
\label{g_in}
\ee
for $r<R$, and consider slices of constant $u$ both inside and outside
the shell.
Then the volume over which one integrates the Einstein-Hilbert
action would have null boundaries at $u=u_i$ and $u=u_f$ for all
$0<r<R_\infty$ (this would change the shape of $\Omega_{in}$ with
respect to Fig.~\ref{mixed}).
It is however easy to show that this does not change the effective
action.
In fact the scalar curvature for Schwarzschild and Minkowski can be
formally written as in Eq.~(\ref{Rv}), with $a_{in}$ a function
of (at most) $r$, since the mass aspect is constant and equal to
$M_{in}$ for $r<R$.
Therefore both the volume contribution and surface terms
vanish in the region $r<R$ and one is left with the terms at
$\eta=-\epsilon$ in $S_G^{shell}$ as given in Eq.~(\ref{SGsh}).
\setcounter{equation}{0}
\section{Curvature scalar for the Vaidya metric}
\label{scalar}
The scalar curvature for a metric of the form (\ref{vaidya})
or (\ref{g_in}) with a generic $a=a(u,r)$ can be computed
straightforwardly from the definition of the Riemann tensor
\cite{mtw}
\be
{\cal R}=g^{\nu\sigma}\,{\cal R}^\mu_{\ \nu\mu\sigma}
\ ,
\ee
where $\mu,\nu\ldots=u,r,\theta,\phi$, and
\be
{\cal R}^\mu_{\ \nu\mu\sigma}=\Gamma^\mu_{\nu\sigma,\mu}
-\Gamma^\mu_{\nu\mu,\sigma}
+\Gamma^\alpha_{\nu\sigma}\,\Gamma^\mu_{\alpha\mu}
-\Gamma^\mu_{\nu\alpha}\,\Gamma^\alpha_{\sigma\mu}
\ .
\ee
The non-vanishing connection coefficients,
\be
\Gamma^\mu_{\nu\lambda}={1\over 2}\,g^{\mu\sigma}\,
\left(g_{\sigma\nu,\lambda}+g_{\sigma\lambda,\nu}
-g_{\nu\lambda,\sigma}\right)
\ ,
\ee
are given by
\be
\begin{array}{lll}
\Gamma^u_{uu}=-{1\over 2}\,{\partial a\over\partial r}\ ,
&
\Gamma^u_{\phi\phi}=
\sin^2\theta\,\Gamma^u_{\theta\theta}=r^2\,\sin^2\theta\ ,
&
\\
& & \\
\Gamma^r_{uu}={1\over 2}\,{\partial a\over\partial u}
+{a\over 2}\,{\partial a\over\partial r}\ ,
&
\Gamma^r_{\phi\phi}=
\sin^2\theta\,\Gamma^r_{\theta\theta}=-a\,r\,\sin^2\theta\ ,
&
\Gamma^r_{ur}={1\over 2}\,{\partial a\over\partial r}\ ,
\\
& & \\
\Gamma^\theta_{r\theta}=\Gamma^\phi_{r\phi}
={1\over r}\ ,
&
\Gamma^\theta_{\phi\phi}=-\sin^2\theta\,\Gamma^\phi_{\theta\phi}
=-\sin\theta\,\cos\theta\ ,
&
\end{array}
\ee
and lead to the result stated in Eq.~(\ref{Rv}).
\setcounter{equation}{0}
\section{Gaussian normal coordinates in Vaidya space-time}
\label{gauss}
The metric in a neighborhood of the shell can be written
in Gaussian normal coordinates $(\tau,\eta,\theta,\phi)$ as
\be
ds^2=g_{\tau\tau}\,d\tau^2+g_{\eta\eta}\,d\eta^2
+2\,g_{\tau\eta}\,d\tau\,d\eta+r^2\,d\Omega^2
\ ,
\label{v_gn}
\ee
where
\be
&&g_{\tau\tau}(\tau,\eta)=-a\,\left({\partial u\over\partial\tau}\right)^2
-2\,{\partial u\over\partial\tau}\,
{\partial r\over\partial\tau}
\nonumber \\
&&g_{\tau\eta}(\tau,\eta)=0
=-a\,{\partial u\over\partial\eta}\,{\partial u\over\partial\tau}
-{\partial u\over\partial\tau}\,{\partial r\over\partial\eta}
-{\partial u\over\partial\eta}\,{\partial r\over\partial\tau}
\nonumber \\
&&g_{\eta\eta}(\tau,\eta)=1
=-a\,\left({\partial u\over\partial\eta}\right)^2
-2\,{\partial u\over\partial\eta}\,{\partial r\over\partial\eta}
\ ,
\ee
and we choose $\eta>0$ ($\eta<0$) for points in $\Omega_{out}$
($\Omega_{in}$).
\par
Matching (\ref{v_gn}) to the three-metric (\ref{3_g}) on the shell
world-volume at $\eta=0$ gives the set of equations
\be
&&A\,\dot U^2+2\,\dot U\,\dot R=N^2
\label{00}
 \\
&&A\,\dot U\,\left.{\partial u\over\partial\eta}\right|_{\eta=0}
+\dot U\,\left.{\partial r\over\partial\eta}\right|_{\eta=0}
+\dot R\,\left.{\partial u\over\partial\eta}\right|_{\eta=0}
=0
\label{01} \\
&&A\,\left(\left.{\partial u\over\partial\eta}\right|_{\eta=0}\right)^2
+2\,\left.{\partial u\over\partial\eta}\right|_{\eta=0}\,
\left.{\partial r\over\partial\eta}\right|_{\eta=0}
=-1
\ .
\label{02}
\ee
From (\ref{00}) it follows that $\dot U=(\pm\beta-\dot R)/ A$,
with $\beta$ given in Eq.~(\ref{beta}).
We require $\dot U>0$ and, since we wish to describe collapsing
trajectories with $\dot R<0$ for $R>2\,M$, we must choose the
plus sign
\be
\dot U={\beta-\dot R\over A}
\ .
\label{dotU}
\ee
For $R\to 2\,M$ the above expression diverges (unless $\beta=\dot R=0$),
which signals the fact that the coordinates in use become singular
on that surface \cite{lind}.
\par
In fact, one has that $u\to +\infty$ for $r\to 2\,m$ (both for Vaidya
and Schwarzschild).
This can be cured from the onset by passing to Israel's coordinates
$(v,w)$ which are regular all the way down to $r=0$ \cite{israel2}.
They are defined by
\be
du=-{dv\over W(v)}
\ ,\ \ \ \ \ \
{dW\over dv}={1\over 4\,m(v)}
\
\ee
and $r=2\,m(v)+W(v)\,w$.
In this frame the Vaidya line element becomes
\be
ds^2=\left({4\over W}\,{dm\over dv}+{w^2\over2\,m\,r}\right)\,dv^2
+2\,dv\,dw+r^2\,d\Omega^2
\ .
\ee
However the same result for the components of the extrinsic curvature
is obtained if the change is made at the end of the computation
(see Appendix~\ref{junction}).
\par
Eqs.~(\ref{01}) and (\ref{02}) yield
\be
\left.{\partial u\over\partial\eta}\right|_{\eta=0}=
-{\beta-\dot R\over A\,N}
\ ,\ \ \ \ \ \ \
\left.{\partial r\over\partial\eta}\right|_{\eta=0}=
{\beta\over N}
\ .
\label{d_eta}
\ee
Hence the radial coordinate $r$ as a function of $\tau$ and
$\eta$ is continuous, but its derivative with respect to $\eta$ has a
jump on $\Sigma$.
At the same time $u$ need not even be continuous across $\Sigma$.
We can now write $(u,r)$ as functions of the Gaussian normal coordinates
$(\tau,\eta)$ explicitly up to order $\eta$,
\be
\left\{\begin{array}{l}
u=U-\strut\displaystyle{\beta-\dot R\over A\,N}\,\eta+{\cal O}(\eta^2)
\\
 \\
r=R+\strut\displaystyle{\beta\over N}\,\eta+{\cal O}(\eta^2)
\ ,
\end{array}\right.
\label{ur}
\ee
where we recall that the above expressions hold both in $\Omega_{in}$
($\eta<0$) and $\Omega_{out}$ ($\eta>0$), thus it is to be understood
that
\be
A=\left\{\begin{array}{ll}
\lim\limits_{\eta\to 0^-} a_{in}=1-{2\,M_{in}/ R}
&\ \ \ \  {\rm in}\ \ \Omega_{in}
\\
\lim\limits_{\eta\to 0^+} a_{out}=1-{2\,M_{out}/ R}
 &\ \ \ \  {\rm in}\ \ \Omega_{out}
\ ,
\end{array}\right.
\ee
and so forth.
\par
From the knowledge of the mapping (\ref{ur}) one can determine the
components (\ref{v_gn}) of the metric and their derivatives with respect
to $\eta$ and compute the extrinsic curvature according to Eq.~(\ref{k}).
\setcounter{equation}{0}
\section{Standard boundary term at $R_\infty$}
\label{border}
Surface contributions are usually computed by making use of the trace
${\cal K}$ of the extrinsic curvature of the border of the space-time
volume $\Omega$, according to Eq.~(\ref{EH}).
Further, ${\cal K}$ is related to the covariant
derivative of the unit normal to the border.
Thus, at $r=R_\infty$, one has
\be
S_B^{(3)}={2\over\kappa}\,\int_{u_i}^{u_f} du\,d\theta\,d\phi\,
\left.\sqrt{h}\,\nabla_\mu\,n^\mu\right|_{R_\infty}
\ ,
\ee
where $h=a\,r^4\,\sin^2\theta$ is the determinant of the
pull-back of the metric (\ref{vaidya}) on an hypersurface of constant
$r$, $\nabla$ denotes the covariant derivative in the metric (\ref{vaidya}),
\be
\nabla_\mu\,n^\mu={1\over\sqrt{-g}}\,\partial_\mu\left(\sqrt{-g}\,n^\mu\right)
\ ,
\ee
and $n^\mu=(-a^{-1/2},a^{1/2},0,0)$ is the unit normal to
an hypersurface of constant $r$ in $(u,r)$ components.
Computing the derivative at $r=R_\infty$ one finds
\be
S_B^{(3)}={8\,\pi\over\kappa}\,\int_{u_i}^{u_f} du\,
\left[2\,R_\infty-{R_\infty^2\over R_\infty-2\,M_{out}}\,{dM_{out}\over du}
-3\,M_{out}\right]
\ .
\label{sb}
\ee
\par
We may now show that Eq.~(\ref{sb}) is dynamically equivalent to the
result (\ref{SB}) used in the text.
We first integrate the second term in the integral above and obtain
\be
S_B^{(3)}={8\,\pi\over\kappa}\,\int_{u_i}^{u_f} du\,
\left[2\,R_\infty-3\,M_{out}\right]
+{4\,\pi\over\kappa}\,R_\infty^2\,\ln{R_\infty-M_f\over R_\infty-M_i}
\ .
\label{SB3}
\ee
The last term does not affect the equations of motion, since the effective
action is varied with fixed endpoints, and can therefore be dropped.
It is shown in Section~\ref{action} that $S^{(2)}_B$ depends on two arbitrary
coefficients and for $\alpha=1$ and $\gamma=0$ one just recovers
the integral in Eq.~(\ref{SB3}).
\section{Junction equations}
\setcounter{equation}{0}
\label{junction}
The junction equations at the shell surface $\Sigma$ relate the
jump in the extrinsic curvature $K_{ij}$ to the matter stress-energy
tensor on $\Sigma$ \cite{mtw},
\begin{eqnarray}
\left[{\cal K}_{ij}\right]^{out}_{in}
=-{\kappa\over 2}\,\left({\cal S}_{ij}
-{1\over 2}\,g_{ij}\,{\cal S}^k_{\ k}\right)
\ ,
\label{junct}
\end{eqnarray}
where $g_{ij}$ is the three-metric (\ref{3_g}) in the proper time
gauge $N=1$ and the source term ${\cal S}_{ij}$ is given in
Eq.~(\ref{S}).
Upon substituting the components (\ref{Kthth}) of the extrinsic curvature
into Eq.~(\ref{junct}) one obtains the $\theta\theta$-junction
equation
\be
E={16\,\pi\over\kappa}\,R\,\left[\beta\right]^{in}_{out}
\label{match1}
\ ,
\ee
where we have introduced the shell proper mass $E$ according to
Eq.~(\ref{EF}).
Analogously (\ref{Ktt}) yield the $\tau\tau$-junction equation
\be
P={2\over\kappa}\,\left\{
\left[{\beta\over R}+{\ddot R\over\beta}+
{M\over R^2\,\beta}\right]^{in}_{out}
+{\dot M_{out}\over\beta_{out}}\,{\beta_{out}-\dot R\over R-2\,M_{out}}
\right\}
\ .
\label{match2}
\end{eqnarray}
In Israel's coordinates (see Appendix~\ref{gauss}) the last term in
the right hand side of Eq.~(\ref{match2}) becomes
\be
{\dot U\,\dot M_{out}\over R\,\beta_{out}}
={1\over R\,\beta_{out}}\,{\dot V^2\over W}\,{dM_{out}\over dv}
\ ,
\ee
which renders our equation (\ref{match2}) the same as Eq.~(3.3) of
Ref.~\cite{pim} where the problem of integrating Eqs.~(\ref{match1})
and (\ref{match2}) was addressed numerically for a particular choice of
the luminosity.
\par
Let us focus on analytic results that can be obtained
independently of the detail of the interaction between
the shell and the out-flowing null dust.
We take, for $\Omega_{in}$, flat Minkowski space ($M_{in}=0$)
and $2\,M_{out}(\tau_i)<R_i$.
\par\noindent
{\bf 1})
Taking $R\gg 2\,M_{out}$ and $\dot R\ll 1$ in Eq.~(\ref{match1})
gives, to leading order, $E\simeq (16\,\pi/\kappa)\,M_{out}$,
which is the relation one would expect in (asymptotically) flat space.
\par\noindent
{\bf 2})
Expanding Eqs.~(\ref{match1}) and (\ref{dH}) for
$R-2\,M_{out}\ll 2\,M_{out}$ gives, again to leading order,
\begin{eqnarray}
&&E\simeq
2\,M_{out}\,\left(\sqrt{1+\dot R^2}+\dot R\right)
\label{RH}
\\
&&\dot E\simeq
-{64\,\pi\over\kappa}\,{M_{out}\,\dot M_{out}\,\dot R\over R-2\,M_{out}}
-8\,\pi\,M_{out}\,\dot R\,P
\ .
\label{RHd}
\end{eqnarray}
For $\dot E=0$ it follows that $\dot M_{out}=0$ only if $P$ always
remains zero (dust).
One can interpret this result as if, in general, the radiation can
be fed by the gravitational energy of the shell whose extraction induces
a surface tension (and changes the motion of $R$ correspondingly).
\par
If we demand that $P$ be regular everywhere, because of the denominator
on the right hand side of Eq.~(\ref{RHd}), for $\dot M_{out}<0$ the proper
energy emitted per unit proper time diverges with $R\to 2\,M_{out}$.
This, together with Eq.~(\ref{RH}), namely $E\sim M_{out}$, implies that
$E$ and $M_{out}$ would vanish at a value of $R\ge 2\,M_{out}$.
On the other hand, if the time derivative of the mass aspect vanishes
according to the bound in Eq.~(\ref{finite}) one (temporarily) recovers
the non-radiating case, for which the shell crosses $2\,M_{out}$ at a
finite proper time and with finite proper energy.
The condition (\ref{finite}) is necessary for the stress-energy tensor
of the radiation to be locally finite when the shell crosses the surface
$r=2\,M_{out}$ \cite{christensen}, that is
\be
\lim\limits_{r\to 2\,M_{out}}
{|T_{uu}^{rad}|\over (r-2\,M_{out})^2}\sim
\lim\limits_{R\to 2\,M_{out}}
{|\dot M_{out}|\over (R-2\,M_{out})}<\infty
\ .
\ee
We remark that the stronger condition $\dot M_{out}=0$ is necessary
to ensure that $r=2\,M_{out}$ is a null surface after the shell
has crossed it, which is a property of event horizons as opposite to
apparent horizons \cite{he}.
In fact, the tangent to the surface $r=2\,M_{out}$ in $\Omega_{in}$
is $t^\mu=(\dot T_{in},2\,\dot M_{out},0,0)$ and has norm
$t^\mu\,t_\mu=-1$.
In $\Omega_{out}$ the tangent would be
$t^\mu\sim(\dot U,2\,\dot M_{out},0,0)$ with norm
$t^\mu\,t_\mu\sim -4\,\dot M_{out}\,\dot U$.
Therefore $t^\mu$ is time-like in $\Omega_{in}$ and would become
space-like (or null for $2\,\dot M_{out}=0$) in $\Omega_{out}$
\cite{lind}.
\par
The mathematical origin of the bound (\ref{finite}) lies in the inadequacy
of the coordinates $(u,r)$ to cover both the interior and the exterior
of $r=2\,m$ in the Vaidya space-time (see Appendix~\ref{gauss}).
It has also a physical interpretation in terms of the point of view
of a static observer at $r\gg 2\,M_{out}$.
On assuming that the metric for $r\gg 2\,M_{out}$ can be approximated by
the Schwarzschild line element, the time $t_\infty$ of the static observer
is related to $\tau$ according to \cite{mtw,guth}
\begin{eqnarray}
{dt_\infty\over d\tau}={\beta\over 1-2\,M_{out}/R}
\simeq -{2\,M_{out}\,\dot R\over R-2\,M_{out}}
\ .
\end{eqnarray}
Then Eq.~(\ref{RHd}) can be written as
\begin{eqnarray}
{dE\over dt_\infty}\simeq {32\,\pi\over\kappa}\,{dM_{out}\over d\tau}
\ .
\label{dE}
\end{eqnarray}
This is in accordance with the fact that, due to the infinite redshift
experienced by a distant observer, the luminosity of a star which
collapses and forms a black hole would decay exponentially and eventually
(for $t_\infty\to+\infty$) fade.
On the other hand, if the flux of radiation measured by the distant
observer does not vanish before all the proper energy of the shell has
been radiated away, then the shell remains always outside the surface
$r=2\,M_{out}$ until $E$ vanishes (thus leading to flat Minkowski space
instead of a black hole).

\begin{thebibliography}{99}
%
\bibitem{mtw}
C.W. Misner, K.S Thorne and J.A. Wheeler,
{\it Gravitation}, W.H. Freeman and Co., San Francisco
(1973).
%
\bibitem{oppenheimer}
J. R. Oppenheimer and H. Snyder, {\em Phys. Rev.} {\bf 56}, 455 (1939).
%
\bibitem{bv}
R. Brout and G. Venturi, Phys. Rev. D {\bf 39}, 2436 (1989).
%
\bibitem{bfv}
C. Bertoni, F. Finelli and G. Venturi, {\em Class. Quantum Grav.}
{\bf 13}, 2375 (1996).
%
\bibitem{ebo}
R. Casadio, {\em Quantum gravitational fluctuations and the semi-classical
limit}, preprint gr-qc/9810073.
%
%
\bibitem{choptuik}
M. Choptuik, {\em Phys. Rev. Lett.} {\bf 70}, 9 (1993).
%
\bibitem{tolman}
R. C. Tolman, {\em Proc. Nat. Acad. Sci.} {\bf 20}, 169 (1934).
%
\bibitem{acvv}
G. L. Alberghi, R. Casadio, G. P. Vacca and G. Venturi,
{\em Class. Quantum Grav.} {\bf 16}, 131 (1999).
%
\bibitem{israel}
W. Israel, {\em Nuovo Cimento} {\bf B 44}, 1 (1966);
{\em Nuovo Cimento} {\bf B 48}, 463 (1966).
%
\bibitem{hk}
P. Hajicek and J. Kijowski, {\em Phys. Rev.} D {\bf 57}, 914 (1998).
%
\bibitem{lake}
K. Lake, {\em Phys. Rev.} D {\bf 19}, 2847 (1979).
%
\bibitem{guth}
E. Farhi, A. H. Guth and J. Guven, {\em Nucl. Phys.} {\bf B 339}, 417 (1990).
%
\bibitem{pim}
R. Pim and K. Lake, {\em Phys. Rev.} D {\bf 31}, 233 (1985).
%
\bibitem{vaidya}
P. C. Vaidya, {\em Proc. Indian Acad. Sci.} {\bf A33}, 264 (1951).
%
\bibitem{israel2}
W. Israel, {\em Phys. Lett.} {\bf 24A}, 184 (1967);
{\em Phys. Rev.} {\bf 143}, 1016 (1966).
%
\bibitem{bicak}
J. Bicak and K. Kuchar, {\em Phys. Rev.} D {\bf 56}, 4878 (1997).
%
\bibitem{h1}
P. Hajicek, {\em Phys. Rev.} D {\bf 57}, 936 (1998).
%
\bibitem{adm}
R. Arnowitt, S. Deser and C. W. Misner, in {\em Gravitation:
an introduction to current research}, L. Witten editor, Wiley,
New York (1962)
%
\bibitem{kuchar}
K. V. Kuchar, {\em Phys. Rev.} D {\bf 50}, 3961 (1994).
%
\bibitem{balbinot}
S. Ansoldi, A. Aurilia, R. Balbinot and E. Spallucci,
{\em Class. Quantum Grav.} {\bf 14}, 2727 (1997).
%
\bibitem{h2}
P. Hajicek, {\em Phys. Rev.} D {\bf 58}, 084005 (1998).
%
\bibitem{bp}
J. Bicak, private communication.
%
\bibitem{Kuc}
The same approach is used in K. V. Kuchar,
{\em Black hole formation by canonical dynamics
of gravitating shells: an equatorial view}, preprint and private
communication.
%
\bibitem{sud}
E. C. G. Sudarshan and N. Mukunda, {\em Classical dynamics: a modern
perspective}, R. Klieger Pub. Co., Malabur, Florida (1983).
%
\bibitem{lind}
R. W. Lindquist, R. A. Schwartz and C. W. Misner, {\em Phys. Rev.}
{\bf 137} (1964) 1364.
%
\bibitem{hawking}
S. W. Hawking, {\em Nature} {\bf 248}, 30 (1974);
{\em Comm. Math. Phys.} {\bf 43}, 199 (1975).
%
\bibitem{york}
Indeed our results are analogous to those obtained for a black hole
in a spherical cavity of radius $R$, see
J. W. York, {\em Phys. Rev.} D {\bf 33}, 2092 (1986).
%
\bibitem{X}
We denote exact differentials by $d$ and generic variations by
$\dbar$.
%
\bibitem{zemansky}
M. W. Zemansky, {\em Heat and thermodynamics}, MacGraw-Hill
Book Company, New York (1968).
%
\bibitem{fpq}
The general solution to Eq.~(\ref{Teq}) (having the units of a temperature)
is given by a sum of powers of $M_{out}$ and the fundamental constants
of the form
\be
{\cal T}_{p,q}={q\over 8\,\pi\,k_b}\,
\left({16\,\pi\,\hbar\over\kappa}\right)^{(1+p)/2}
\,\left({16\,\pi\,M_{out}\over\kappa}\right)^{-p}
\,{1\over\sqrt{1-2\,M_{out}/R}}
\ ,
\nonumber
\ee
and for $p=q=1$ one obtains the temperature (\ref{Tyork}) one would
expect a distance $R$ from a black hole of mass $M_{out}$ \cite{york}.
%
\bibitem{christensen}
S. M. Christensen and S. A. Fulling,
{\em Phys. Rev.} D {\bf 15}, 2088 (1977).
%
\bibitem{he}
S. W. Hawking and G. F. R. Ellis,
{\em The large scale structure of space-time},
Cambridge University Press, Cambridge (1973).
%
%
%
\end{thebibliography}
\end{document}